\documentclass[aps,twocolumn,groupedaddress,showpacs]{revtex4}
\usepackage{amsmath}
\usepackage{amssymb}
\usepackage{amsfonts}
\usepackage[dvips]{graphicx}
\usepackage[]{caption}
\usepackage[]{epsfig}
\begin{document}

\title{Electric field-controlled water permeation coupled to ion transport through a nanopore}

\author{J. Dzubiella}
\email[e-mail address: ] {jd319@cam.ac.uk}
\affiliation{University Chemical Laboratory,
Lensfield Road,
Cambridge CB2 1EW,
United Kingdom}
\author{R. J. Allen}
 \affiliation{University Chemical Laboratory,
Lensfield Road,
Cambridge CB2 1EW,
United Kingdom}

\author{J.-P. Hansen}
 \affiliation{University Chemical Laboratory,
Lensfield Road,
Cambridge CB2 1EW,
United Kingdom}

\date{\today}

\begin{abstract}
We report Molecular Dynamics (MD) simulations of a generic
hydrophobic nanopore connecting two reservoirs which are initially at
different Na${\bf{^+}}$ concentrations, as in a biological cell. The nanopore is
impermeable to water under equilibrium conditions, but the strong electric field
caused by the ionic concentration gradient drives water molecules
in. The density and structure of water in the 
pore are highly field dependent. In a typical simulation run, we
observe a succession of cation passages through the pore,
characterized by approximately bulk  mobility. These ion passages
reduce the electric field, until the pore empties of water and closes
to further ion transport, thus providing a possible mechanism for
biological ion channel gating.   
\end{abstract}


\maketitle

Water and ion permeation of nanopores is a key issue for
biological membrane-spanning ion channels and aquaporins, as well as for
materials like zeolites, gels and carbon nanotubes. Recent simulations
report intermittent filling of hydrophobic nanopores by water under
equilibrium conditions \cite{hummer:nature,beckstein:jpc,
  allen:prl}. However, imbalances in ion 
concentrations in the inside and outside of  cell membranes create
strong electric fields \cite{hille}. Experiments on water
near interfaces show that strong fields can induce considerable 
electrostriction of water \cite{toney:nature}. The nonequilibrium
behavior of confined water and ions in strong fields should therefore
be very important for ion permeation and ion channel function. In
particular, biological ion channels ``open'' and ``close'' to ion
transport in response to changes in the electric field across the
membrane. This behavior, know as {\it voltage gating}, is crucial to
their function, but its mechanism is so far not well understood
\cite{hille}. Related approaches \cite{kuyucak:2001} to ion transport include
the application of a uniform external electric field \cite{crozier:prl,chung:2002}, more
specific models of particular proteins
\cite{berneche:nature,tieleman:2001,roux:1994,lopez:2002}, Brownian
Dynamics \cite{chung:2002} and
continuum theories \cite{nonner:1998}. Here we present the results of MD simulations in which a
strong electric field across the pore is explicitly created by an ionic
charge imbalance, as in a cell. We follow the relaxation of this
nonequilibrium system to equilibrium.

Our generic model ion channel consists of a cylindrical hydrophobic pore of
length $L_{\rm p}= 16$\AA~ and radius $R_{\rm p}=5$ to $7$\AA, through a membrane
slab which separates two reservoirs containing water and Na$^+$ and Cl$^-$
ions, as shown in Fig. 1. One reservoir has initial
concentrations $c_{\rm {Na^{+}}}\approx 0.9$M (12 cations) and $c_{\rm
  Cl^{-}}\approx 0.6$M (8 anions), while for the second reservoir, $c_{\rm 
  Na^{+}}\approx 0.3$M (4 cations) and $c_{\rm Cl^{-}}\approx 0.6$M (8
anions). This imbalance of charge generates an average electric field
of $0.37$V/\AA~ across the membrane. These ion concentrations and
electric field are typically five times larger than under ''normal``
physiological conditions, but they could be achieved in the course of a
rare, large fluctuation at the pore entrance. The enhanced
ion concentrations in the initial state were chosen to improve the
signal-to-noise ratio in the simulation and to allow the detection of
novel transport mechanisms.
The chosen pore dimensions are 
comparable to those of the selectivity filter of a K$^{+}$  channel \cite{zhou:nature,berneche:nature}.

The simulation cell contains two pores in sequence along the z-axis,
one of which is shown in Fig.1. The reservoir to the right of this
pore thus forms the left reservoir for the other pore. Due to
periodic boundary conditions, the right reservoir of the latter
is also the left reservoir of the first. In this arrangement, the
simplest to allow the use of full three-dimensional periodic
boundaries, the flows of ions in response to the concentration
gradient will be anti-parallel in the two channels. The relaxation
towards equilibrium, where the two reservoirs are individually
electroneutral, will thus involve an indirect coupling between the two
pores. 

The water molecules are modelled by the SPC/E potential 
\cite{berendsen:jpc} which consists 
of an O atom, carrying an electric charge
$q=-0.8476$e, and two H atoms with $q=0.4238$e. The O-atoms on
different water molecules interact via a Lennard-Jones (LJ) potential
with parameters $\epsilon=0.6502$kJmol$^{-1}$ and  $\sigma=3.169$\AA~. The
model is rigid, 
with OH bond length $1$\AA~ and HOH angle $109.5^\circ$. The Na$^+$
ions have LJ parameters $\epsilon=0.3592$kJmol$^{-1}$, $\sigma=2.73$\AA~ and $q=+$e
and for the Cl$^-$, $\epsilon=0.1686$kJmol$^{-1}$, $\sigma=4.86$\AA~ and
$q=-$e \cite{spohr:1999}. 

Ions and water O-atoms interact with the confining pore and membrane
surfaces by a potential of the LJ form
$V=\epsilon'[(\sigma'/r)^{12}-(\sigma'/r)^{6}]$, where $r$ is the
orthogonal distance from the nearest confining surface. The potential
parameters are $\epsilon'=1.0211$kJmol$^{-1}~$ and
$\sigma'=0.83$\AA~. If $R_{\rm p}$ is the geometric radius of the
cylindrical pore, one may conveniently define an effective radius $R$
by the radial distance from the cylinder axis at which the interaction
energy of a water O-atom with the confining surface is $k_{B}T$,
leading at room temperature to $R\approx R_{\rm p}-2$\AA~; similarly the
effective length of the pore is $L\approx L_{\rm p}+4$\AA~. Ion-water,
ion-ion, water-surface and ion-surface cross terms are defined using
the usual Lorentz-Berthelot combining rules. 
Polarizability of
the membrane \cite{allen:jpcm} and of the water molecules and ions is
neglected.     

The total simulation cell including both of the channels is of
dimensions $l_{x}=l_{y}=23.5\pm 0.3$\AA~ and 
$l_{z}=112.9\pm1.5$\AA~  and contains 1374
water molecules, 16 Na$^+$ and 16 Cl$^-$ ions. Molecular Dynamics
simulations were carried out with the DLPOLY2 package \cite{dlpoly},
using the Verlet algorithm \cite{frenkelsmit,allen} with a timestep of 2fs. The
pressure was maintained at $P=1$bar and the temperature at $T=300$K using
a Berendsen barostat and thermostat \cite{berendsen:jcp}. Electrostatic
interactions were calculated using the particle-mesh Ewald
method \cite{essmann:jcp}.

  \begin{figure}[htb]
  \begin{center}
    \epsfig{file=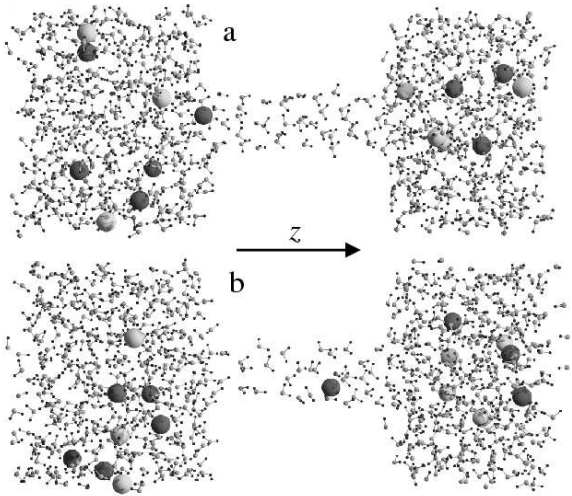, width=8cm, angle=0}
    \caption{Simulation snapshots. Molecular configurations (a) before  a cation
   (dark grey spheres) permeates the channel and (b), 10 ps later,  while it goes
   through. Anions are shown as light grey spheres. Only half of the
   periodically repeated simulation cell is shown.}
  \label{profiles}
\end{center}
\end{figure}

  \begin{figure}[htb]
  \begin{center}
    \epsfig{file=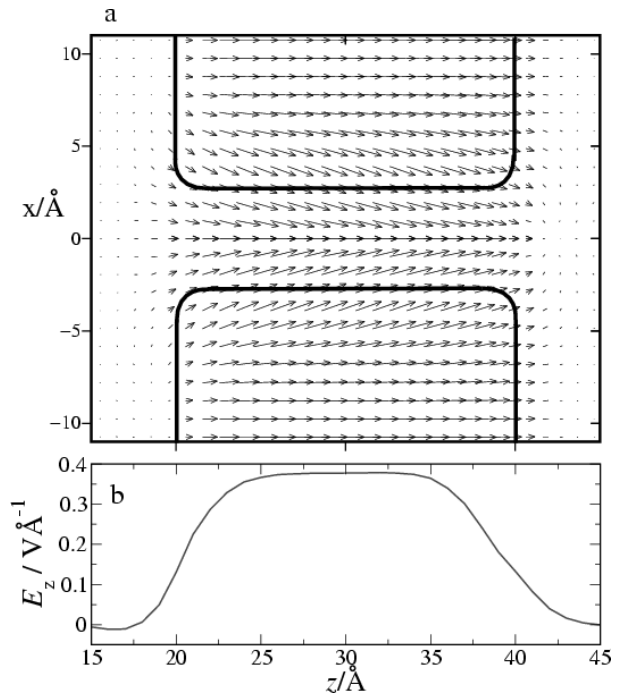, width=8cm, angle=0}
 \caption{Electric field in and around the pore. (a), Magnitude
   and direction of the electric field, depicted by vectors on a
   two dimensional rectangular grid in the $x$-$z$-plane. The contour
   of the membrane pore is sketched as black solid line. (b),
   Averaged $z$-component $E_{z}(z)$ of the electric field inside the
   channel.}
  \label{profiles}
\end{center}
\end{figure}

  \begin{figure}[htb]
  \begin{center}
    \epsfig{file=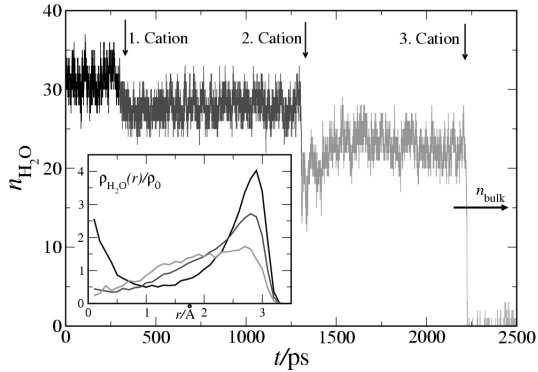, width=6cm, angle=-90}
    \caption{Occupancy and structure of water inside the pore. The number of
   water molecules inside the channel $n_{\rm H_{2}O}$ is plotted as a
   function of time $t$. The 
   shades of grey code the average magnitude of the electric field inside the
   pore: from black to light grey: $E \approx 0.37 V/$\AA, $E \approx 0.25 V/$\AA,
   $E \approx 0.15V/$\AA~, $E \approx 0.10V/$\AA. The inset shows the
   corresponding radial density profiles 
   of the water molecules inside the channel averaged over periods of
   constant field. $\rho_{0}$ is the bulk density of water in
   equilibrium. $n_{\rm bulk}$ (eq. (2)) is the expected number of water
   molecules inside the pore if bulk density is assumed.}
  \label{profiles}
\end{center}
\end{figure}

  \begin{figure}[htb]
  \begin{center}
    \epsfig{file=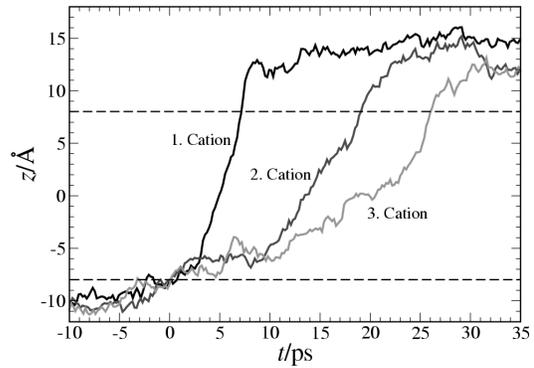, width=6cm, angle=-90}
    \caption{Cation trajectories inside the pore and in its vicinity. The $z$-coordinate of
   the cations is plotted versus time for three successive permeations
   in one typical simulation run. The channel is located between
   $z=-8{\rm\AA}$ and $z=8{\rm\AA}$ marked by the two long dashes
   lines. $t=0$ defines the time at which the ions are located at the
   entrance of the channel. The shades of grey  code the average magnitude of
   electric field experienced by the ions, as in Fig.3}
  \label{profiles}
\end{center}
\end{figure}
Water permeation of the pore is strongly affected by the electric
field. The effective channel radius chosen for
most of the simulations ($R=3$\AA) 
is such that under equilibrium conditions (i.e. with equal numbers of
anions and cations on both sides and hence no electric field), the channel is
empty of water and ions \cite{hille}. However, the ionic charge
imbalance across the the membrane causes the pore to fill
spontaneously with water. The electric field 
throughout the system was monitored by measuring the electrostatic
force on phantom test particles on a three dimensional
grid \cite{crozier:prl}. Fig.~2a shows the average local electric
field around one pore before the first 
ion moves through a channel. It is nearly zero in the reservoirs.
 Inside the pore the field is
very strong ($\sim 0.37$V/\AA) and has a small inward radial component. The profile
of the $z$-component of the field $E_{z}$ is shown in Fig. 2b and is
constant inside the pore.

During
the course of the simulation, a number of Na$^+$ ions move through the
pore. Each of these events changes the reservoir charge imbalance and
reduces the electric field in the pore. This has a dramatic effect on
the behavior of the water, as shown in Fig.~3 in which the number
$n_{\rm H_{2}O}$ of water molecules inside one pore is plotted as a
function of time for a typical simulation run. 

Initially, the water
in the pore undergoes strong electrostriction, comparable to
experimental observations \cite{toney:nature}, with an average density
\begin{equation}\label{mth1}
\rho\approx n_{\rm H_{2}O}/(\pi R^{2}L),
\end{equation}
twice as large as that of bulk water in
equilibrium. If we assume bulk density of water $\rho_{0}$ inside a
channel of 
radius $R$ and length $L$ we expect an average number $n_{\rm bulk}$
of molecules inside the channel with
\begin{equation}\label{mth2}
n_{\rm bulk}=\rho_{0}(\pi R^{2}L).
\end{equation}
$n_{\rm bulk}$ is indicated with an arrow on the right-hand side in
Fig. 3. 
At each ion crossing, the number $n_{\rm H_2O}$ of water
molecules inside the pore drops but is still larger than $n_{\rm
  bulk}$. In this particular simulation run shown in Fig.3 three
cations went through this pore.
After the third ion crosses ($t\approx 2.25$ns), the
electric field is no longer strong enough to sustain channel filling
and the pore spontaneously empties of water, thereby becoming
impermeable to further ion transport. However the other pore in the
simulation cell remains filled and the final ion crossing eventually
occurs through this pore which after that also empties of
water. Finally equilibrium is restored to the system and both channels
are empty of water. On repeating the simulation 5 times we observe
that the closing of one pore after the third ion passed through it
occurs in all runs.  

The structure of water in the filled pore is strongly affected by the field, as
shown in the inset of Fig.~3.
 Before the first ion crossing, water forms clear layers near
the pore wall and along the z-axis. The central layer disappears after
the first ion crossing and the outer layer becomes less
well-defined. After a further ion crossing, water is more-or-less
evenly distributed. 

Ion transport through the pore is found to occur essentially at
constant velocity. Fig. 1 
shows snapshots from a typical simulation run just before (a) and
while (b) a sodium ion passes through the channel. Within the
reservoirs the anions and cations diffuse among the water
molecules. When a cation in the Na$^{+}$-rich reservoir comes 
close to the channel entrance, it experiences the strong axial field
shown in Fig. 2 and is dragged into the channel. Analysis of $15$
simulation runs, with the same initial charge imbalance but 
different initial configurations, shows that once the first ion enters
the channel, it moves with a constant velocity which is approximately
the same in all runs, and then reverts to diffusive motion at the
other end of the channel. 

Fig. 4 shows typical cation positions along
the z-axis, as a function of time, for the first, second and third ion
crossings, as shown in Fig. 3. The second ion also traverses the
channel at constant, somewhat reduced velocity, although it appears to
pause for approximately 10ps at the channel entrance, perhaps in order
to shed its bulk-like solvation shell. We observe that this ``pausing
time'' is rather widely distributed between simulation runs. 
The cation mobility $\mu_+$, defined by $\vec
v = \mu_{+} e \vec E$, can be calculated from the slopes of the
trajectories in Fig. 4, together with the measured electric fields, as in
Fig.2. The resulting values are $\mu_{+}\approx 4.5\times
10^{11}{\rm\;s\,kg^{-1}}$ for the first ion, $\mu_{+}\approx 3.8\times
10^{11}{\rm\;s\,kg^{-1}}$  for the second ion, and  $\mu_{+}\approx 2.4\times
10^{11}{\rm\;s\,kg^{-1}}$ for the third ion. These values are close to
the value of $\mu_{+}\approx2.3\times 10^{11}{\rm\;s\,kg^{-1}}$
obtained from the self-diffusion constant in the reservoir $D_+$,
using Einstein's relation $\mu_{+}=D_{+}/k_{B}T$, but seem to increase
with the magnitude of the electric field inside the pore. This
enhancement of the mobility correlates with the change of 
structure of the water inside the channel, shown in the inset of Fig. 3. 
The tetrahedral hydrogen bond network which water forms under
equilibrium conditions is disrupted inside the pore under high
electric fields.

Simulations of wider ($R=5$\AA) pores and with different lengths give
qualitatively the same results. The critical electric
field for water permeation is, however, sensitive to the pore radius
and length. This suggests that voltage-dependent gating in ion
channels, if it were to occur by changes in water permeation of a
hydrophobic section of the pore \cite{allen:prl,hille}, might be strongly
dependent on channel geometry.    

The key finding which emerges from our simulations is the strong
correlation between water and ion behavior under non-equilibrium
conditions. Ionic charge imbalance across the membrane induces water
permeation of the hydrophobic pore and thus makes it permeable to
ions. This suggests that voltage gating of ion channels may be linked to the
coupling between water and ion permeation in pores far from
equilibrium. The structure and density of water in the pore is
dramatically 
affected by the strong electric field.  The passage of a cation
through the channel causes an abrupt jump of the electric 
field, and an ensuing jump in the number of water molecules inside the
pore. Ion passage through the pore occurs at constant velocity and
with a mobility coefficient similar to that of the bulk solution at
equilibrium. 

\acknowledgments 
The authors are grateful to
  Jane Clarke and Michele Vendruscolo for a careful reading of the
  manuscript. This work was supported in part by the EPSRC. R.~J.~A. is grateful
  to Unilever for a Case award.

\end{document}